\documentclass[graybox,envcountchap]{svmult}

\usepackage{mathptmx}        
\usepackage{amsmath}
\usepackage{amssymb}
\usepackage{color}
\usepackage{helvet}          
\usepackage{courier}         
\usepackage{dirtree}

\usepackage{makeidx}        
\usepackage{graphicx}        
\usepackage{subfig}

\usepackage{multicol}        
\usepackage[bottom]{footmisc}

\usepackage[numbers]{natbib}
\usepackage{hyperref}        
\hypersetup{colorlinks=true,urlcolor=blue}

\usepackage[misc]{ifsym}

\makeindex             

\newcommand{\pd}{\partial}

\newcommand{\apj}{Astrophys.\ J.\ }
\newcommand{\apjl}{Astrophys.\ J.\ }
\newcommand{\apjs}{Astrophys.\ J. Suppl.\ }
\newcommand{\aap}{Astron.\ Astrophys.\ }

\newcommand{\azh}{Astron.~Zh.\ }

\newcommand{\prd}{Phys.\ Rev.\ D\ }
\newcommand{\pre}{Phys.\ Rev.\ E\ }
\newcommand{\prl}{Phys.\ Rev.\ Lett.\ }
\newcommand{\araa}{Annu.\ Rev.\ Astron. Astrophys.\ }

\newcommand{\nat}{Nature~}
\newcommand{\physrep}{Phys.\ Rep.\ }
\newcommand{\mnras}{Mon.\ Not.\ R.\ Astron.\  Soc.\ }

\newcommand{\apss}{Astrophys.\ Space Sci.\ }

\newcommand{\pasa}{Publ.\ Astron.\ Soc.\ Australia}

\newcommand{\baas}{Bull.\ Am.\ Astron.\ Soc.\ }

\newcommand{\ud}{\mathrm{d}}

\begin{document}

\title{Supernova Simulations}

\author{Bernhard M\"uller}

\institute{Bernhard M\"uller (\Letter) \at School of Physics and Astronomy, Monash University, 10 College Walk, Clayton, VIC~3800,  Australia, \email{bernhard.mueller@monash.edu}}

\maketitle

\abstract{Magnetohydrodynamic simulations of core-collapse supernovae have become increasingly mature and important in recent years. Magnetic fields take center stage in scenarios for explaining hypernova explosions, but are now also considered in supernova theory more broadly as an important factor even in neutrino-driven explosions, especially in the context of neutron star birth properties. Here we present an overview of simulation approaches currently used for magnetohydrodynamic supernova simulations and sketch essential physical concepts for understanding the role of magnetic fields in supernovae of slowly or rapidly rotating massive stars. We review  progress on simulations of neutrino-driven supernovae, magnetorotational supernovae, and the relevant field amplification processes. Recent results on the nucleosynthesis and gravitational wave emission from magnetorotational supernovae are also discussed. We highlight efforts to provide better initial conditions for magnetohydrodynamic supernova models by simulating short phases of the progenitor evolution in 3D to address uncertainties in the treatment of rotation and magnetic fields in current stellar evolution models.}

\section{Introduction}
\label{sec:1}
Just like during the later lives of neutron stars and accreting black holes that were discussed in the preceding chapters, magnetic fields play an important role in the  formation for these relativistic compact objects by stellar collapse. By far the most important formation channel is the collapse of massive stars that have proceeded through advanced nuclear burning stages to form an iron core that then becomes unstable to collapse due to electron captures and photodisintegration of heavy nuclei \citep{bethe_90,shapiro_83}. In most instances, the collapse of the iron core also results in an explosion that ejects the outer layers of the star as a \emph{core-collapse supernova}.\footnote{Note that this term is sometimes applied more broadly to the entire compact object formation event, even when no explosion occurs.}
While the association of observed core-collapse supernovae with compact remnants \citep{hewish_68,holland_84} and with massive progenitor stars \citep{smartt_09b,smartt_15} is now solidly established, the explosion mechanism has proved more difficult to unravel as electromagnetic observations cannot directly probe the supernova core. While neutrinos \citep{mueller_19d} and gravitational waves \citep{kalogera_19,abdikamalov_22} may in future supply direct information on the supernova engine, simulations remain an essential tool for understanding the inner working of core-collapse supernovae.

Magnetic fields were identified as a crucial piece of the puzzle early on in the quest for the core-collapse supernova explosion mechanism. In the early history of the field, the extraction of rotational energy from the collapsed core was seriously considered as a generic supernova mechanism
\citep{bisnovatyi_70,leblanc_70,meier_76,bisnovatyi_76}, but a more nuanced and tempered picture of the role of magnetic fields in core-collapse supernovae has emerged since then. Current understanding of stellar rotation points to rather slow core rotation rates in massive stars \citep{heger_05,ma_19} and is therefore not compatible with magnetorotational powering as a generic supernova mechanism. Explosions driven by rotation and magnetic fields, either in the context of rapidly spinning ``milliseconds magnetars''
\citep{uzov_92,duncan_92} or black holes with accretion disks (collapsars) \citep{macfadyen_99,macfadyen_01} remain the most promising scenario for  hyperenergetic supernovae (``hypernovae'', observationally classified as broad-lined Ic supernovae) \citep{woosley_06b}, which account for about 1\% of supernovae in the local Universe.

This, however, has only made the study of magnetohydrodynamics (MHD) in core-collapse supernovae much more nuanced and certainly no less relevant. Three-dimensional (3D) simulations of MHD-driven explosions are now readily available and inching towards a meaningful comparison, and there are incipient efforts to better connect them to the intricate problem of angular momentum transport and magnetic field amplification in stellar evolution. The diversity of extreme supernovae from classical hypernovae to superluminous supernovae opens up the possibility of a greater variety of scenarios for MHD-powered explosions. Furthermore, it is increasingly recognized that magnetic fields probably play an important role in supernovae with ordinary explosion energies as well, albeit not as the primary driver of the explosion. Perhaps most importantly, MHD effects in core-collapse supernovae are key to the understanding of neutron star birth spin periods and magnetic fields.

\begin{figure}
    \centering
    \includegraphics[width=\linewidth]{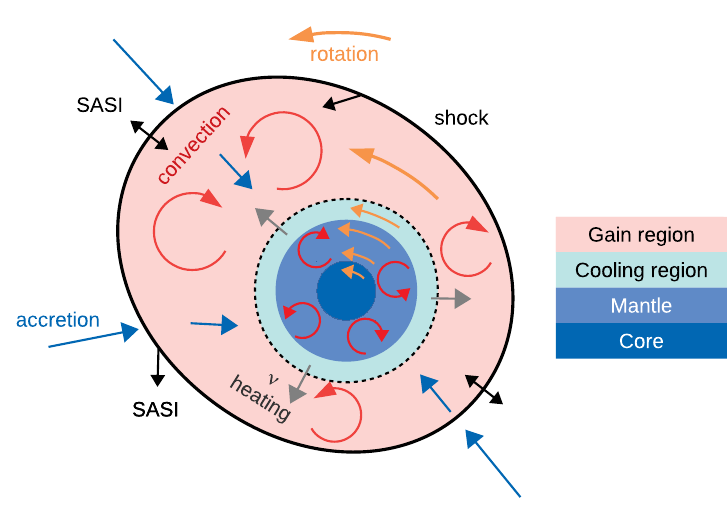}
    \caption{Sketch of the structure of the supernova core during the post-bounce phase before the onset of explosion, top view along the rotation axis. The
    proto-neutron star consists of a non-convective core,
    a convective mantle, and a convectively stable cooling 
    layer at the surface (progressively lighter shades of blue).     
    Neutrinos emitted from the proto-neutron star (gray) deposit
    energy in the gain region (pink) behind the shock (black),
    and can also drive convective overturn there. 
    Note also the accretion flow (blue arrows) onto the proto-neutron star, which is supersonic outside and subsonic inside the shock. In addition to
    convective flow in the gain region and the proto-neutron star (indicated by red circular arrows), the shock may be unstable to SASI oscillations
    (black arrows). Differential rotation may be present throughout
    the supernova core (orange arrow). Convection, SASI, differential rotation and shear flow in and near the cooling region can drive amplification of magnetic fields individually or in concert.
    }
    \label{fig:supernova_core}
\end{figure}

\section{Progenitor Structure and Post-Collapse Structure of the Supernova Core}
It is useful to first outline the conditions in the progenitor star and in the core of the supernova after
collapse to provide essential background to the reader. This will also serve to organize our discussion of MHD in core-collapse supernovae by charting the conditions pertinent to the evolution and impact of magnetic fields in different regions of the stellar core for different pathways of collapse and explosion. 

\subsection{Conditions in the Progenitor Star}
\label{sec:prog}
Many of the conditions for the magnetohydrodynamic evolution during the supernova are already set prior to the collapse of the iron core. The onion shell structure of the massive progenitor star, specifically the pre-collapse density profiles, determine the accretion rate onto the young proto-neutron star after the collapse of the iron core, which in turn affects the ram pressure onto the supernova shock and the accretion power that can be converted into neutrinos and/or magnetic fields. For perfectly spherical and non-rotating stellar models, the pre-collapse density structure is therefore the primary determinant for whether or not an explosion will occur or not, and for explosion and remnant properties \citep{ertl_15,mueller_16a}. Real stars will generally rotate, will exhibit asphericities, e.g., in convective regions \citep{couch_15,mueller_16b,yadav_19,mueller_20,fields_21}, and will contain seed magnetic fields.
For the magnetohydrodynamic evolution after collapse, the seed fields are relevant because they may jump-start the amplification of magnetic fields in the supernova core, though  they likely become less relevant on longer time scales, when field amplification processes in the young neutron star and its vicinity have had sufficient time to reach saturation levels. The angular momentum distribution of the progenitor influences the rotational energy of the young neutron star after collapse, the degree of differential rotation within it and around surface, and the possibility of disk formation due to centrifugal support. It therefore impacts magnetic field amplification by differential rotation after collapse in the neutron star and its environment, and the rotational energy reservoirs that magnetic fields can tap and potentially convert into outflows.

Except for tentative ventures into 3D stellar hydrodynamics (see Section~\ref{sec:precollapse}), magnetic fields and rotation currently need to be modeled using effective recipes for magnetic field amplification and angular momentum transport in one-dimensional (1D) stellar evolution models. This introduces considerable quantitative uncertainties, especially since current 1D stellar evolution models struggle to fit constraints on interior rotation stellar profiles from asteroseismology \citep{cantiello_14,belkacem_15,eggenberger_17}. Naturally, the pre-collapse field geometry remains a major unknown, even though features of the field geometry can sometimes be deduced from the nature of the underlying amplification mechanisms. For example, field amplification by differential rotation via the Tayler-Spruit dynamo \citep{spruit_02} is expected to produce predominantly toroidal fields. 

The current paradigm of stellar evolution \citep{heger_05,fuller_19,ma_19},
which is informed by measurements for low-mass post-main sequence stars and white-dwarfs, suggests that core spin rates of massive stars are typically in the range of tens to hundreds of seconds \citep{heger_05},  and that core magnetic fields are of order $10^9\,\mathrm{G}$ with predominantly toroidal components. Rare evolutionary scenarios that avoid strong core spin-down 
and achieve pre-collapse core rotation rates of $\mathord{\sim} 1\, \mathrm{s}$ can also be accommodated \citep{aguilera_20}. These would translate into millisecond-period neutron star spin periods after collapse without angular momentum redistribution and possibly provide conditions for various flavor of MHD-driven explosions for hypernovae. However, such rapid rotation rates cannot be plausibly combined with pre-collapse magnetic fields stronger than $10^9\texttt{-}10^{10}\,\mathrm{G}$ because higher field strengths would come at the expense of more efficient core spin-down. In addition, there may be a route towards very slow pre-collapse rotation and unusually strong magnetic fields. Very strong fields can be generated in stellar mergers, and if these can be retained as fossil fields up to core collapse, central field strengths
$10^{11}\texttt{-}10^{12}\,\mathrm{G}$
may be reached \citep{schneider_19,schneider_20}. After some inevitable reconnection, such merger-generated fields are expected to settle to a twisted-torus configuration whose decay is inhibited due approximate conservation of magnetic helicity \citep{braithwaite_06}.

\subsection{Structure of the Supernova Core after Collapse}
Eventually the iron core of the progenitor star becomes unstable to gravitational collapse due to deleptonisation and photo-disintegration of heavy nuclei \citep{shapiro_83,bethe_90}. The collapse proceeds beyond nuclear saturation density and is eventually halted by the stiffening of the nuclear equation of state, which leads to a rebound (``bounce'') of the inner core, and launches a shock wave into the surrounding shells that are still collapsing supersonically. Even for rapid progenitor rotation, centrifugal forces cannot stop the collapse below nuclear densities \citep{dimmelmeier_07_a}.

Due to the dissociation of infalling heavy nuclei in the infalling matter and rapid neutrino emission in the
neutrino burst, the shock weakens and turns into an accretion shock within milliseconds \citep{bethe_90,rampp_02,liebendoerfer_04}.  
What emerges in the supernova core is a layered structure with
a high-density ``proto-neutron star'' at the center, an accretion shock that is initially driven to a radius about $100\texttt{-}200\,\mathrm{km}$, and a subsonic accretion flow from the supernova shock down to the proto-neutron star (Figure~\ref{fig:supernova_core}).
Some mechanism for extracting energy from the proto-neutron star and/or the accretion flow is required to ``revive'' the shock and initiate an explosion. The two currently favored scenarios are energy deposition by neutrinos from the proto-neutron star and its surface 
(neutrino-driven mechanism) or
the extraction of rotational energy from the proto-neutron star by magnetic fields (magnetorotational mechanism), see \citep{bethe_90,janka_12,mueller_20} for reviews.

The proto-neutron star itself consists of a low-entropy, high-density core, a hotter, roughly adiabatically stratified mantle of densities around $10^{14}\, \mathrm{g}\,\mathrm{cm}^{-3}$, and an atmosphere (surface) with a
steep positive entropy gradient and negative density gradient in the region where neutrinos decouple from matter and cool the proto-neutron star surface. The region between the proto-neutron star and the shock is roughly adiabatically stratified, and will later experience net heating by neutrinos (at which point it is termed \emph{heating} or \emph{gain region}). This stratification of the supernova core region is critical for magnetohydrodynamic effects because it determines the nature of instabilities and flow patterns that will dictate magnetic field amplification. The proto-neutron star mantle and the gain region tend to permit instabilities that involve radial fluid flow (e.g., convection, \citep{herant_94,burrows_95,janka_96}), whereas the proto-neutron star surface and the core-mantle interface are convectively stable, but may permit magnetic field amplification by shear flows. With minor modifications, this overall structure still persists initially after the shock is revived and starts to move outward.
Accretion downflows and neutrino-driven outflows probably coexist in the gain region for several seconds \citep{mueller_15b,mueller_17,bollig_21}, fueling fresh material down to the bottom of the gain region for efficient heating and then transporting the hot neutrino-heated material outward to energize the explosion. Eventually, accretion will subside, and a neutrino-driven wind outflow from the proto-neutron star will develop \citep{duncan_86}, although the onset of a ``true'' spherical wind is much delayed in recent simulations \citep{mueller_15b,mueller_17,bollig_21,wang_23}.

With our focus on magnetic field effects in core-collapse supernovae, the rotational state of the supernova core and the magnetic field configuration after collapse are of critical importance. As a first crude approximation, one can assume that the collapse is perfectly spherical, in which case the amplification of magnetic fields during collapse by compression can be obtained easily from flux conservation. The radial magnetic field component $B_r$ after collapse is given in terms of the initial field strength $B_{r,0}$ and
the initial and final radial coordinates $r_0$ and
$r$ of the fluid element under consideration as
\begin{equation}
    B_r=B_{r,0}  \left(\frac{r_0}{r}\right)^2.
\end{equation}
For the meridional and azimuthal field components
$B_\theta$ and $B_\varphi$, magnetic flux conservation
leads to a different relation to the pre-collapse
values $B_{\theta,0}$ and $B_{\varphi,0}$,
\begin{eqnarray}
    B_\theta&=&B_{\theta,0}  \left(\frac{r_0}{r}\right)\left(\frac{\pd r}{\pd r_0}\right)^{-1},\\
    B_\varphi&=&B_{\varphi,0}  \left(\frac{r_0}{r}\right)\left(\frac{\pd r}{r_0}\right)^{-1}.\\
\end{eqnarray}
If the collapse is homologous ($r \propto r_0$), 
the result is the same as for $B_r$. Homologous (self-similar) collapse \citep{goldreich_80,yahil_83} holds very well for the inner part of
the iron core out to mass coordinate of
$\sim 0.5 M_\odot$ for no or slow rotation \citep{langanke_03}, and the scaling of the magnetic field with $r^{-2}$ remains a useful zeroth-order approximation in other contexts as well. Flux conservation alone would imply that for progenitor fields of 
$10^{9}\texttt{-}10^{10}\,\mathrm{G}$, the collapse of an iron core of $\mathord{\sim} 1000\,\mathrm{km}$ in radius to a neutron star of $\mathord{\sim} 12\,\mathrm{km}$,
fields of order $ \lesssim 10^{14}\, \mathrm{G}$ will be reached in the neutron star. This, however, refers to the field strength at the \emph{center} of the \emph{final}, cold neutron star. The warm proto-neutron star in the supernova core maintains several times its
final radius for several hundreds of milliseconds after bounce so that the surface field strengths that can be reached by compression during the collapse are about two orders of magnitude lower.

For the evolution of the angular velocity $\omega$ during (spherical) collapse, angular momentum conservation also leads to a scaling with
$r^{-2}$,
\begin{equation}
    \omega \approx \omega_0 \left(\frac{r_0^2}{r}\right)^{2},
\end{equation}
in terms of the pre-collapse angular velocity $\omega_0$ of a fluid element. This leads to an interesting corollary. Differential rotation can be a key driver for magnetic field amplification, but magnetic fields in turn tend to counteract differential rotation, which has long led to the notion that the core and the surrounding burning shells in massive stars are close to rigid rotation (for an extensive discussion, see \citep{mcneill_22}), although they may rotate differentially with respect to each other. Under
the assumption of spherically symmetric contraction, the rate of differential rotation after collapse is given by
\begin{equation}
    \frac{\pd \omega}{\pd r}
    =\frac{\pd r^{-2} j(r)}{\pd r}
    =\frac{\pd r^{-2} (r_0^2 \omega_0(r))}{\pd r}
    =
    \frac{r_0^2 \pd \omega_0/\pd r_0}{r^2 \pd r/\pd r_0}+\frac{2 r_0 \omega_0}{r^2 \pd r/\pd r_0}-\frac{2 r_0^2 \omega_0}{r^3},
\end{equation}
where $j$ denotes the specific angular momentum of a fluid element.
This implies that differential rotation ($\pd \omega /\pd r \neq 0$) will emerge in regions that collapse non-homologously even from shells that rotate uniformly in the progenitor. Shear-driven instabilities such as the magnetorotational instability (MRI, \citep{balbus_91,balbus_98,akiyama_03}) can then occur in such regions.

\section{Specific Requirements for Core-Collapse Supernova Simulations}
Simulating the dynamics of magnetic fields in the supernova core poses different challenges than the environments discussed in previous chapters. Here we outline a few salient points and refer to in-depth reviews \citep{mueller_20,mezzacappa_20} for details. Supernova explosions, especially neutrino-driven ones, are in a sense ``slow'' phenomena that develop and unfold over many dynamical time scales. Explosions can take hundreds of milliseconds to develop and seconds to reach their final energies \citep{mueller_15b,mueller_17,bollig_21,burrows_24}. The accretion flow behind the shock is moderately subsonic, and convective flow in the proto-neutron star is very subsonic with Mach numbers of order of $\mathcal{O}(10^{-2})$. This imposes stringent requirements on the ability of codes to maintain hydrodstatic stability, conserve the total energy (ADM mass in general relativity) for a self-gravitating fluid, and to conserve angular momentum, especially in scenarios where rotation is dynamically important; all of this may require special tweaks \citep{mueller_20}. Accurate neutrino transport is critical for the dynamics, for multi-messenger observables, and for accurate predictions of supernova nucleosynthesis. Various neutrino transport algorithms are currently in use
\citep{mezzacappa_20}, but multi-group neutrino transport, detailed interaction rates, and at least an approximate treatment of general relativistic transport effects (in particular gravitational redshift) is required. Supernova modeling is a multi-scale problem; structures of several hundred meters in the proto-neutron star need to be captured at the same time as shells in the star at radii of at least $10,000\texttt{-}100,000\,\mathrm{km}$. The self-gravity
of the stellar core and the proto-neutron star is crucial in the supernova explosion problem.  This necessitates a dynamical treatment of gravity, though the approximation of a fixed space-time metric may be viable for simulating phenomena like wind outflows from the proto-neutron star or hydrodynamics instabilities in the gain region to some degree in idealized simulations.

Oidealized simulationsn the other hand, newly formed proto-neutron stars are less compact than cold ones, and inflow and outflow velocities are at best mildly relativistic well into the explosion phase -- even in the case of magnetorotational supernovae -- which often permits a pseudo-Newtonian treatment of relativistic gravity by means of a modified gravitational potential \citep{marek_06,mueller_08}. However, more extreme conditions are encountered in black-hole forming collapse and during later explosion stages when the environment of the nascent compact objects becomes more dilute and relativistic outflows can develop. Special relativistic hydrodynamics with a (pseudo-)Newtonian approximation is sometimes used as an intermediate step towards general relativistic MHD \citep{takiwaki_09}. The merit of these approaches is that they can provide a better trade-off of physical and numerical model accuracy (e.g., better neutrino transport or full-wave Riemann solvers), computational cost and fidelity in treating relativistic effects.

A variety of schemes for the MHD equations have been in use for core-collapse supernova simulations.
The codes \textsc{3DnSNe-IDSA} \citep{takiwaki_16,matsumoto_20},
\textsc{Aenus-Alcar} \citep{obergaulinger_14},
\textsc{CoCoNuT-FMT} \citep{mueller_15a,mueller_20b},
\textsc{fGR1} \citep{kuroda_20},
\textsc{FISH} \cite{kaeppeli_11},
\textsc{Flash-M1} \citep{oconnor_18a},
and
\textsc{GRHydro} (part of the Einstein toolkit; 
\citep{moesta_14a}) are capable of 3D MHD supernova simulations. Of these, 
\textsc{fGR1} and  \textsc{GRHydro} are fully general relativistic, whereas the other codes resort to the Newtonian approximation with an effective potential; \textsc{Aenus-Alcar} 
is also capable of special relativistic MHD.
Constrained transport for enforcing the divergence-free condition \citep{evans_88} is used in
\textsc{Aenus-Alcar},  \textsc{fGR1}, \textsc{FISH}, and \textsc{GRHydro}, whereas
\textsc{3DnSNe-IDSA}  and \textsc{CoCoNuT-FMT} implement slightly different flavors of hyperbolic divergence cleaning \citep{dedner_02} (which is also available in \textsc{GRHydro}). Historically, 
constrained transport, hyperbolic divergence cleaning and projection methods have all been used in the \textsc{Flash} code framework
\citep{crockett_05,lee_13,derigs_18}. Older two-dimensional (2D) codes for supernova modeling have used schemes similar in spirit to constrained transport build the divergence-free condition directly into the grid discretization  \citep{livne_07,ardeljan_05}.

Higher-order reconstruction schemes are used in all modern MHD supernova codes. One virtue of Newtonian codes like \textsc{3DnSNe-IDSA},
\textsc{Aenus-Alcar}, \textsc{CoCoNuT-FMT}, and  \textsc{Flash-M1} lies in greater ease in implementing less diffusive Riemann solvers like HLLC \citep{toro_94} or HLLD \citep{miyoshi_05} with superior performance in the subsonic and sub-Alfv\'enic regime, whereas general relativistic supernova simulations have so far been limited to the more diffusive HLLE \citep{harten_83} solver. A particular subtlety in MHD simulations concerns the difficulties of proper upwinding \citep{gardiner_05}, which may make some implementations prone to (possibly mild) instabilities under certain conditions. The most ambitious and rigorous approaches for proper upwinding involve reconstruction of characteristic variables \citep{gardiner_05,lee_13,stone_20}, which is not currently implemented in any MHD supernova code except
\textsc{Flash} \citep{lee_13}.
For specialized purposes, e.g., for the long-term evolution of very subsonic convection in the proto-neutron star over time scales of seconds, other tailored numerical approaches are used, such as spectral methods in combination with the anelastic approximation \citep{raynaud_20}.

The repercussions of these differences in the implementation of the MHD equations, and those of other approximations (pseudo-Newtonian approximation, different neutrino transport treatments) need to be gauged by code comparisons for verification and uncertainty quantification.
A number of supernova code comparisons have been performed \citep[e.g.,][]{liebendoerfer_05,mueller_10,mueller_15a,oconnor_18c,just_18,cabezon_18}, but there is only one MHD code comparison in 2D between \textsc{Aenus-Alcar} and \textsc{CoCoNuT-FMT} so far \citep{varma_21b}.

\section{Supernovae of Non-Rotating and Slowly Rotating Stars}
While magnetic fields in core-collapse supernovae have historically been studied predominantly in the context of rapidly rotating progenitors, it is useful to first consider magnetic fields in the more generic case of non-rotating or slowly rotating progenitors. Although magnetic fields will almost invariably play at most a subsidiary role to neutrino heating in powering the explosion, the inclusion of magnetic fields is becoming important for precision modeling and indispensable for eventually understanding the origin of neutron star birth magnetic fields and spins.

The growth and potential dynamical role of magnetic fields in supernova cores in the absence of rapid rotation was already discussed before the advent of detailed multi-dimensional simulations.
\citet{thompson_93} already pointed to the possibility of turbulent dynamo action in the proto-neutron star due to convective instability
and estimated saturation field strength.
It has also been theorized that Alfv\'en waves emanating from the proto-neutron star
\citep{suzuki_08}
and Alfv\'en wave amplification within the gain region \citep{guilet_11} may play a role the supernova explosion mechanism and for supernova nucleosynthesis
during the neutrino-driven wind phase \citep{suzuki_05}. 

Since around 2010 the amplification of magnetic fields in supernova cores by flow instabilities not reliant on progenitor rotation has been studied extensively in multi-D simulations. Two such instabilities are known to operate in the gain region. First, neutrino heating makes the accretion flow behind the shock onto the proto-neutron star unstable to convection
\citep{herant_94,burrows_95,janka_96}. Second, the accretion shock is unstable to large-scale oscillatory motions, typically dominated by
the dipole ($\ell=1$) component, known as
standing-accretion shock instability (SASI).
SASI takes on the form of sloshing motions \citep{blondin_03}
or in 3D also of circular motion (spiral mode) \citep{blondin_07}, and is mediated by an amplification cycle involving vortical and acoustic perturbations in the region behind the shock and the neutron star surface \citep{foglizzo_07,guilet_12}.
Idealised 2D MHD simulations of SASI without neutrino transport showed the potential for amplification of magnetic fields by
about two orders of magnitude, but without a significant dynamical effect
\citep{endeve_10}. Two-dimensional simulations with neutrino transport \citep{obergaulinger_14} also suggested limited potential for field amplification by convection and SASI by only a factor of a few; the key limitation for field amplification in these simulations was the short time for stretching and distorting magnetic fields by convective motions before they are advected out of the gain layer onto the neutron star surface. The simulations of \citet{obergaulinger_14} raised the interesting prospect of a dynamical effect of magnetic fields on supernova explosions in the case of very strong pre-collapse fields of order $10^{12}\,\mathrm{G}$, which may be encountered for certain types of supernova progenitors, e.g., those arising from stellar mergers (see Section~\ref{sec:prog}).
A similar supporting role for very strong magnetic
fields was also found in later 2D simulations
with neutrino transport \citep{matsumoto_20,jardine_22}.

\subsection{Impact of Magnetic Fields on Shock Revival in 3D Simulations}
A principal limitation of these early works lay in the restriction to axisymmetry, which precludes the operation of magnetic dynamos \citep{cowling_33}. In core-collapse supernovae of non-rotating or slowly-rotating progenitors, at least small-scale turbulent dynamo amplification is expected to occur in the gain region (regardless of whether SASI or convection is the dominant instability) and in the proto-neutron star convection zone. The small-scale dynamo operates without the need for bulk differential rotation \citep{brandenburg_05}, driven by turbulent 
flow that may arise, e.g., due to convection, SASI, or shear instabilities. Turbulent dynamos are typically characterized by fast exponential growth with a growth time scale of the order of the eddy turnover time \citep{schober_12}.
The initial exponential growth phase typically slows down and may transition to linear growth once the
magnetic fields approach kinetic equipartition 
for higher wavenumbers \citep{cho_09}, so that backreaction of the magnetic field on the flow becomes important. At low Reynolds number, the small-scale dynamo requires a magnetic Prandtl number (the ratio of kinematic viscosity and magnetic resistivity)
$\mathrm{Pr}_\mathrm{m}\gtrsim 1$ 
\citep{haugen_04,schober_12}, a condition which is easily fulfilled in the gain region. There are, however, still debates in dynamo theory about possible adverse effects on small-scale dynamo amplification and its saturation in the limits of high Reynolds and high-magnetic Prandtl number
\citep{schekochihin_02,stepanov_08,schober_15}. By and large, both simulations \citep{warnecke_23}
and astrophysical observations \citep{christensen_09,brun_17} point to a higher robustness of the small-scale dynamo, however. Small-scale dynamo amplification may in fact become possible well below  $\mathrm{Pr}_\mathrm{m}=1$ in high-Reynolds number flow \citep{warnecke_23}. Observational phenomenology interestingly supports the relatively ``naive'' notion that saturation field strengths close to kinetic equipartition are reached \citep{christensen_09,brun_17}. In interpreting MHD supernova simulations, it is important to consider this context from dynamo theory and astrophysical simulations in addition to raw simulation results; one should not take for granted that explosion models (which invariably employ the ideal MHD approximation) reproduce the correct physical limit in the supernova core.

The possibility of dynamo amplification in the collapse of non-rotating progenitors was first investigated by means of idealized adiabatic 3D simulations of the SASI \citep{endeve_12}. These simulations showed rapid amplification of magnetic fields throughout the post-shock region, fed by vorticity that is generated by shear flows in the spiral mode of the SASI. Despite amplification of magnetic fields to about 10\% of equipartition, the simulations of \citet{endeve_12} did not reveal any dynamical impact of magnetic fields on the shock dynamics.

\begin{figure}
    \centering
\includegraphics[width=0.65\linewidth]{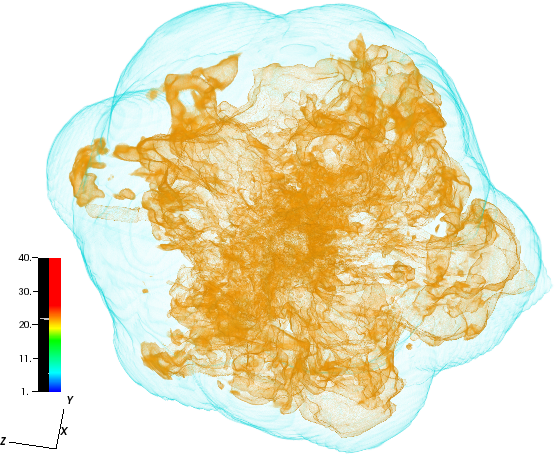}
\\
\includegraphics[width=0.65\linewidth]{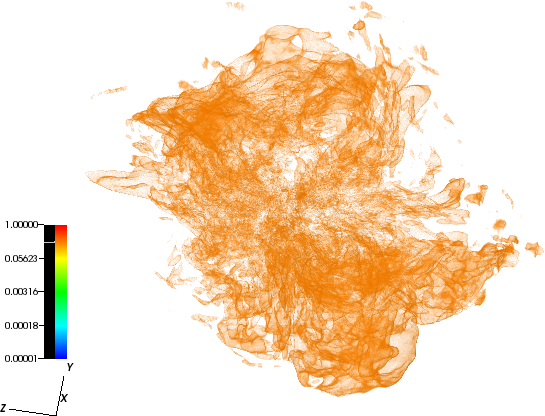}
    \caption{Volume rendering of the entropy
    (top) and the ratio of magnetic to gas pressure
    (bottom) in a high-resolution
    ($0.7^\circ$ in angle) MHD supernova simulation of a non-rotating $20 M_\odot$ stripped-envelope progenitor at a post-bounce time of $250\,\mathrm{ms}$. At this time, an explosion is already underway.  
    In addition to neutrino-heated high-entropy bubbles (orange) and regions of strong fields (orange), the shock is visible a light blue translucent surface in the top panel. The magnetic field configuration is dominated by small-scale structures. Fields of relatively homogeneous strength are present throughout the post-shock region, but they are strongest in and in between the expanding neutrino-heated bubbles. The entire structure has a diameter of about $1200\,\mathrm{km}$.
     }
    \label{fig:mhd_slowrot}
\end{figure}

The first 3D MHD supernova simulation with multi-group neutrino transport for a ``generic'', slowly-rotating $15 M_\odot$ progenitor star accorded a more relevant role to turbulent dynamo amplification \citep{mueller_20b}. With a conservative choice of weak pre-collapse field strengths
of $\mathord{\lesssim} 10^7\,\mathrm{G}$,
\citet{mueller_20b} find relatively rapid small-scale dynamo amplification of the fields in the gain region to about $40\%$ of kinetic equipartition within about $250\,\mathrm{ms}$ after bounce. Figure~\ref{fig:mhd_slowrot}
shows the emerging structure of neutrino-heated convective bubbles and dynamo-generated magnetic fields from a recent high-resolution simulation of a $20 M_\odot$ star with the \textsc{CoCoNuT-FMT} code.

The magnetic fields appear to act as an extra energy reservoir in the gain region, i.e., they do not reduce the turbulent kinetic energy compared to a non-magnetic baseline simulation. \citet{mueller_20b} find that magnetic fields support shock revival, tilting the balance in favor of explosion, whereas no explosion develops in the non-magnetic baseline model.
A similar behavior --- higher combined kinetic and magnetic turbulent energy and more precipitous explosions with magnetic fields --- was seen in more recent supernova simulations of $20 M_\odot$
and $27 M_\odot$ stars with the \textsc{3DnSNe-IDSA} code \citep{matsumoto_22,matsumoto_24}.
It should be noted, however, that magnetic fields play a subsidiary role to neutrino heating and aspherical mass motions in triggering an explosion. It is still neutrino energy deposition that powers the explosion; magnetic fields merely play an auxiliary, but noticeable role in pushing the stalled accretion shock out a bit to the point that neutrino heating can initiate continued shock expansion as a runaway process according to the classical neutrino-driven paradigm \citep{janka_01}. In fact, the ratio of magnetic to kinetic turbulent energy drops again after shock revival in the simulation of \citep{mueller_20b}.

Naturally, these recent simulations prompt questions about the potential impact of resolution and the accuracy to which the implemented schemes for ideal MHD approximate the correct physical limit of high hydrodynamic and magnetic Reynolds number (see Figure~\ref{fig:mhd_slowrot}
for a snapshot from a high-resolution simulation
at $0.7^\circ$ with the
\textsc{CoCoNuT-FMT} code).
Further studies are clearly required to address these issues, but they may to some extent be subordinate to the question of initial conditions.
One implication of these recent simulations with relatively fast turbulent field amplification is that the pre-collapse field strengths may be relevant to post-collapse dynamics in the supernova core. This is particularly interesting for the aforementioned strong fossil-field scenario in stellar merger products. Using a $16 M_\odot$ post-merger stellar evolution model \citep{schneider_20} and a parameterized twisted-torus initial field of about $10^{12}\,\mathrm{G}$, \citet{varma_23a} found a rapidly developing supernova explosion. The rapid explosion in their model appears to come about only in part due to additional magnetic stresses behind the shock, but more so because the magnetic field affects the infall of material onto the shock, giving rise to large-scale ram pressure asymmetries ahead of the shock, which facilitates a strongly unipolar explosion. Intriguingly, early shock revival at a time of high mass accretion rate and high neutrino luminosities provide the conditions for dumping considerable neutrino energy into the explosion, resulting in an explosion energy 
of more than $0.9\times 10^{51}\,\mathrm{erg}$, very close to canonical supernova explosion energies.

\subsection{Pre-Collapse Magnetic Fields from 3D Simulations}
\label{sec:precollapse}

Both for ``normal'' supernova progenitors and those that presumably harbor a strong fossil-field, there is, of course, a need to put the predictions of pre-collapse fields from stellar evolution theory on a more solid footing. In recent years, 3D models of convection and nuclear burning have emerged as a means for more consistently initializing supernova simulations (see \citep{mueller_20} for a review).
This approach has also been extended to MHD, but there are as yet no full-star simulations, and reaching steady-state conditions in stellar magnetoconvection model remains challenging as this would require simulations of very long duration. Pragmatically, a major focus in 3D modeling of stellar convection has been on oxygen shell burning because this shell is typically active at collapse, has relatively high turbulent Mach numbers of order $0.1$ \citep{collins_18}, and is the shell that falls through the shock during the phase when an explosion is likely to start.

Simulations of convective oxygen shell burning provide a first glimpse tat what realistic pre-collapse fields in massive stars may look like \citep{varma_21a}. Field strengths are predicted to saturate well below equipartition with the turbulent kinetic energy, or about $10^{10}\,\mathrm{G}$ in absolute terms. Such fields would not have an important dynamical impact on stellar convection, but may still have an appreciably impact \emph{after} collapse by jump-starting the turbulent dynamo in the gain region. Importantly, values of order $10^{10}\,\mathrm{G}$ in the oxygen shell cannot be directly compared to the expected core field strengths in current 1D stellar evolution models, as the oxygen shell sits at much lower densities at radii of several thousands of kilometers; values of $10^{10}\,\mathrm{G}$ in the oxygen shell are in fact stronger than what was conventionally expected. However, the fields are predicted to be dominated by small-scale structures, with dipole field strengths about an order of magnitude lower than isotropic field strengths.
Uncertainties about the saturation of the magnetic field strength during the pre-collapse evolution exist in both direction. Due to lower magnetic Prandtl number $\mathrm{Pr}_\mathrm{m}\lesssim 1$, 
high efficiency of the turbulent dynamo and high saturation field strengths are not to be taken for granted, but on the other hand, high-resolution simulations \citep{leidi_23} suggest that saturation field strengths in turbulent convective flow may still be underestimated and actually reach a high fraction of kinetic equipartition.

\subsection{Neutron Star Magnetic Fields and Magnetized Outflows}
One of the major questions in compact object formation concerns the origin and distribution of neutron star magnetic fields. Observed neutron star magnetic fields show a broad distribution between
$10^{12}\,\mathrm{G}$ and
$10^{14}\,\mathrm{G}$ for young pulsars, and
a population of \emph{magnetars} with field strengths above $10^{14}\,\mathrm{G}$ and up to
about  $10^{15}\,\mathrm{G}$ \citep{olausen_14}.
Many evolutionary steps are interposed in the evolution of a proto-neutron star during the first
few seconds of a supernova to an observable pulsar or magnetar. During the later explosion phase, ablation of material from the proto-neutron star surface by the neutrino-driven wind will also affect the neutron star surface fields. Ongoing accretion onto the neutron star early on during the explosion as well as late-time fallback may also have an impact; for example, there is the possibility of field burial below fallback material \citep{payne_04,torres_16}. Magnetic fields generated by a dynamo \emph{inside} the warm young neutron star
(see Section~\ref{sec:amplification}) may break out through the surface.
These complications limit the conclusions that can be drawn on the origin of neutron star magnetic fields from current supernova simulations.

Nonetheless, some important insights can be gleaned already. Field amplification by shear motions at the bottom of the gain region is found to easily generate magnetic fields of order $10^{13}\,\mathrm{G}$ and sometimes even $10^{14}\,\mathrm{G}$ at the proto-neutron star surface in 3D supernova simulations
\citep{mueller_20b,varma_23a}. Dipole field strengths in simulations are somewhat lower, but are still at the upper end of the pulsar magnetic field distribution and sometimes verge into the magnetar regime. Interestingly, memory of the pre-collapse fields appears to be lost quickly, probably due to
a balance between field amplification and turbulent reconnection in the neutron star surface region.
Simulations starting from very weak and
very strong twisted-torus fields of
$10^{6}\,\mathrm{G}$ and $10^{12}\,\mathrm{G}$ \citep{varma_23a}
showed little difference in proto-neutron surface field strengths at the end of the runs. These simulations also eventually show  a decline in surface field strength in the explosion phase. Such a decline may be physical and result from turbulent reconnection and ablation of material from the fluffy surface of the warm neutron star, and could lead to more typical pulsar magnetic fields eventually. However, numerical reconnection may also play a role and calls for future high-resolution long-time MHD supernova simulations of the first seconds of the explosion until accretion onto the neutron star ceases. Over longer time scales, innovative (combinations of) numerical methods will be required to deal with the transition to the very different regime of a neutron-star with a very dilute environment and force-free magnetosphere, and potential complications like fallback.

Some of the phenomena that unfold on longer time-scales have in fact already been addressed in simulations using somewhat idealized setups and initial conditions, and simplifications to relevant physics such as neutrino heating during the neutrino-driven wind phase. The interplay of strong magnetic fields and wind outflows and the resulting effects on neutron star spin-down and nucleosynthesis in the wind are of particular interest \citep{thompson_18,prasanna_22,desai_23}.
In the presence of strong large-scale neutron star surface fields $\gtrsim 10^{14}\,\mathrm{G}$, the neutrino-driven wind not only becomes asymmetrical, but also unsteady with episodic mass ejection in plasmoids \citep{thompson_18}. The altered dynamics of such strongly magnetized winds may allow for the ejection of small amounts of high-entropy material with conditions for rapid neutron capture process (r-process) nucleosynthesis up to the third r-process peak \citep{thompson_18,desai_23}. Simulations of magnetized winds also indicate that the early spin-down of highly-magnetized young neutron stars by such outflows may be much more efficient than the classical dipole spin-down formula\footnote{Using the classical dipole spin-down model, the derivative of the spin period $P$ is given in terms for the surface dipole field $B_\mathrm{surf}$ as $\dot P/P \approx \left(\frac{B_\mathrm{surf}}{3\times 10^{19}\,\mathrm{G}}\right)^2 \,\mathrm{s}^{-1}$, see, e.g., \citep{lorrimer_04}.} 
would suggest \citep{prasanna_22}.

\section{Magnetorotational Explosions}
The magnetorotational explosion mechanism has a long history dating back to the 1970s \citep{leblanc_70,bisnovatyi_70,meier_76}, and its key idea has remained current ever since. The rotational energy of rapidly rotating proto-neutron star provides a significant energy reservoir that can be tapped by magnetic fields to deliver the requisite energy to revive the shock and power an explosion. However, beyond this classical version of the magnetorotational mechanism, there is a number of closely related scenarios that 
involve outflows powered by rotation and magnetic fields in core-collapse supernovae. This also includes explosions powered by outflows from accretion disks after black-hole formation in stellar collapse (\emph{collapsar scenario} \citep{macfadyen_99,macfadyen_01}), and covers mechanisms that power different types of outflows during different phases of a supernova. Rotation and magnetic fields may deliver the explosion energy contained in the bulk of the ejecta during the first seconds of the explosion, but they are also relevant for explaining the relativistic jet of long gamma-ray bursts either from rapidly rotating ``millisecond magnetars'' \citep{duncan_92,uzov_92,komissarov_07,bucciantini_08,bucciantini_09} or collapsars 
\citep{macfadyen_99,macfadyen_01}, and they may power dilute late-time wind outflows that shape the energetics and the electromagnetic emission of the supernova on longer time scales \citep{bucciantini_06,metzger_15}.

Within the confines of this chapter, we shall focus mostly on the classical scenario of explosions driven by neutron star rotation within the first hundreds of milliseconds after bounce, both because it is arguably the best-explored one and because readers may find it easier to build up an end-to-end picture of MHD-powered supernovae step-by-step from the collapse phase rather than skipping ahead to later phases.

\subsection{Basic Physical Considerations}
The starting point for the classical magnetorotational mechanism is the reservoir
of rotational energy that can be tapped by magnetic fields. Taking the neutron star
moment of inertia to be
$I\approx 0.36  M R^2$
(which roughly holds for a neutron star
of $1.4 M_\odot$ and radius $12\,\mathrm{km}$, cp.\ \citep{lattimer_05}), the rotational
energy of a neutron star is 
\begin{equation}
    E_\mathrm{rot}=\frac{1}{2}I \omega^2
    =
    \frac{2 \pi ^2 I}{P^2}
    =2\times 10^{52}\,\mathrm{erg}
    \left(\frac{M}{1.4 M_\odot}\right)
    \left(\frac{R}{12 \,\mathrm{km}}\right)^2
    \left(\frac{P}{1 \,\mathrm{ms}}\right)^{-2},
\end{equation}
in terms of its gravitational mass $M$,
radius $R$, and spin period $P$. For millisecond spin periods,
this provides a sufficiently large energy reservoir to be compatible with the range of observed hypernova energies up to
$\sim 10^{52}\,\mathrm{erg}$
\citep{woosley_06b,mazzali_14} if the rotational energy can be extracted efficiently. However, one needs to bear in mind proto-neutron stars remain warm and inflated for at least a few hundred milliseconds, so for a given amount of angular momentum, the available reservoir of rotational energy will be smaller.

Furthermore, magnetic fields may not be able to tap the entire rotational energy on short time scales. The entire reservoir is available only on such time scales over which the neutron star  efficiently exchanges angular momentum with its environment, i.e., initially with a dense accretion flow and later with an increasingly dilute environment. If the coupling to the environment is slow, one excepts that magnetic fields are at best able
to establish uniform rotation within the neutron star.
In this case, at most the free energy contained in differential rotation can be liberated \citep{dessart_07a}. Self-sustained differential rotation in the neutron star convection zone, which is known from stellar and planetary convection \citep{jermyn_20a,jermyn_20b,mcneill_22}, could limit the available energy even further. In practice,
liberation of free rotational energy can still be sufficient for reaching hypernova energies \citep{grimmett_21}, especially because the neutron star can act as a transducer, accelerating accreted material into corotation and ejecting large amounts of it with little net growth in mass.

The dynamics of a supernova with a rapidly rotating proto-neutron
star therefore hinges on the efficiency of energy and angular momentum extraction by magnetic fields. Extraction can in principle happen both by collimated jet outflows, or by more isotropic, wind-like outflows. Jets can emerge due to the interplay of hoop stresses and ambient pressure that leads to the formation of magnetic towers
\citep{dessart_07a},
but whether this occurs and whether the resulting jets are stable depends on various factors including the poloidal-to-toroidal field ratio or the initial magnetic field geometry (see Section~\ref{sec:mrsims}). At this point, it is sufficient to consider order-of-magnitude estimates for the dynamical role of magnetic fields. Assuming that rotational motions dominate the flow, the angle-integrated (isotropic) radial Poynting flux $F_B$ 
can be obtained from $\mathbf{B} (\mathbf{v}\cdot \mathbf{B})/
4\pi$ term in the MHD energy equation and
is given by
\begin{eqnarray}
F_B&\sim & r^2 v_\mathrm{\varphi} B_r B_\varphi
\sim r^3 \omega B_r B_\varphi
\sim r^3 \omega B^2
\label{eq:eflx}
\\
&
\nonumber
\sim&
1.1\times 10^{52}\,\mathrm{erg}\,\mathrm{s}^{-1}
\left(\frac{\omega/2\pi}{\times 10^3\,\mathrm{s}^{-1}}\right)
\left(\frac{r}{12\,\mathrm{km}}\right)^3
\left(\frac{B_r}{10^{15}\,\mathrm{G}}\right)
\left(\frac{B_\varphi}{10^{15}\,\mathrm{G}}\right)
\end{eqnarray}
in terms of the rotational velocity $v_\varphi$, and
the radial and zonal magnetic field components $B_r$, $B_\varphi$
(all of which are to be interpreted as averages or typical values
at radius $r$). Similar estimates can be made for the angular momentum flux $F_L$ due to magnetic stresses, which is of order
\begin{equation}
    \label{eq:lflx}
    F_L\sim B_r B_\varphi r^3.
\end{equation}
Well inside the proto-neutron star, field strengths in excess of $10^{15}\,\mathrm{G}$ amount to only a minuscule fraction of thermal or kinetic equipartition field strengths, and fast redistribution of rotational and angular momentum towards relatively rigid rotation
is therefore plausible. However, the time scale for extracting rotational energy from the neutron star is less trivial. Fast extraction on the time scale of $\sim 1\,\mathrm{s}$ will require
rotational periods of $\sim 1\,\mathrm{ms}$ and magnetar field
strength $\sim 10^{15}\,\mathrm{G}$ at the base of accretion flow
onto the proto-neutron star surface.

The principal questions in the magnetorotational
paradigm are therefore how fast magnetic fields amplified, what saturation field strengths are reached, and also whether ``ordered'' magnetic fields are generated in the supernova core. Magnetic fields need to be ordered in the sense that $B_r$ and $B_\varphi$ in the above equations must be (strongly) correlated to achieve a
net angular momentum and energy flux mediated by magnetic fields, or else the implied integrals over solid angle
in Equation~(\ref{eq:eflx}) and (\ref{eq:lflx}) would fall far below order-of-magnitude estimates based on dimensional analysis. They also need to be organized in large-scale structures for the formation of jets as described above.

As already discussed for the non-rotating case, dynamically relevant magnetic fields cannot be generated simply by compression during collapse from the pre-collapse fields predicted by currently stellar evolution models. In the rotating case, however, additional dynamo mechanisms driven by rotation come into play. Furthermore, in regions with meridional flow (e.g., in the gain region or
in the proto-neutron star convection zone) on top of rotational flow, substantial non-vanishing kinetic helicity of the flow,
\begin{equation}
    H=\int \mathbf{v} \cdot (\nabla \times \mathbf{v})\,\ud V,
\end{equation}
can enable the formation of large-scale fields by an inverse magnetic helicity cascade
\citep{frisch_75,pouquet_76,brandenburg_05}.

In principle, differential rotation alone could lead to continuous field amplification in the collapsed cores of rapidly rotating progenitors, but the amplification would be linear with time and not sufficient to amplify weak initial fields to dynamically relevant ones on short time scales. Exponentially growing amplification mechanisms are required for this purpose.
Two such mechanisms that are discussed prominently in the context of magnetorotational explosions are the magnetorotational instability, which was first discussed in the context of accretion disks \citep{balbus_91,balbus_98} and later identified as relevant to core-collapse supernovae \citep{akiyama_03}, and the
$\alpha$-$\Omega$ dynamo, whose importance in proto-neutron star convection had been recognized even earlier \citep{duncan_92}.

One interesting feature of the magnetorotational instability (MRI) is that it can be driven by differential rotation alone even in convectively neutral or convectively stable regions. Entropy and lepton number gradients will naturally affect the growth and
character of the MRI, however \citep{obergaulinger_09}. For introductory purposes, it is sufficient to review key features
of the MRI in the absence of buoyancy effects as established in the theory of the MRI in accretion disks
\citep{balbus_91,balbus_98,akiyama_03}. 
In the presence of a negative angular velocity gradient,
the MRI leads to unstable growth of modes down to
a minimum wavelength $\lambda_\mathrm{MRI}$ set by the Alfv\'en velocity
$v_\mathrm{A}$ and the angular velocity $\omega$,
\begin{equation}
\lambda_\mathrm{MRI} \sim v_\mathrm{A}/\omega
\sim \frac{|B|}{\sqrt{4 \pi \rho} \omega}
\sim
450 \,\mathrm{cm}
\left(\frac{B}{10^{12}\,\mathrm{G}}\right)
\left(\frac{\rho}{10^{10}\,\mathrm{g}\,\mathrm{cm}^{-3}}\right)^{-1/2}
\left(\frac{\omega/2\pi}{10^3\,\mathrm{s}^{-1}}\right)^{-1}.
\end{equation}
Modes of shorter wavelength are stabilized by magnetic tension.
The maximum growth rate $\Omega_\mathrm{max}$ occurs for modes close to $\lambda_\mathrm{MRI}$, and is of order
\begin{equation}
\Omega_\mathrm{max}=
1/2 (\ud \ln\omega/ \ud \ln r).
\end{equation}
For larger wavelengths $\lambda$, the growth rate roughly scales as $\lambda^{-1}$. Thus, for moderately strong initial fields and short neutron star spin periods, numerical resolution on the meter-scale would be required to resolve the fastest growing mode. At coarser resolution,
the resolved modes on the grid will grow significantly more slowly.
Simple arguments for saturation can be made essentially based on dimensional arguments by requiring that the critical wavelength
$\lambda_\mathrm{MRI}$ will grow due to increasing field strength
to about the shearing length $\ud r/\ud \ln \omega$ before termination of growth, which results in saturation fields of order \citep{akiyama_03}
\begin{equation}
B^2
\sim 4\pi \rho r^2 {\omega}^2 \frac{d\ln\omega}{d\ln r}.
\end{equation}
However, such simple dimensional arguments may not hold. For example, disruption of channel models by resistive instabilities may provide a saturation mechanism for terminating MRI growth at smaller scales and lower field strength \citep{obergaulinger_09}, which makes the question of MRI saturation in core-collapse supernovae highly non-trivial. Amplification may be limited to 
factors of $10\texttt{-}100$ \citep{rembiasz_16}.

Field amplification by an $\alpha$-$\Omega$ dynamo within the
proto-neutron star convection zone \citep{duncan_92} is interesting in that it provides a natural mechanism for generating large-scale fields due to the inverse helicity cascade. With an initial exponential growth rate of 
\citep{vishniac_05}
\begin{equation}
 \Omega \sim (v_\mathrm{conv}/L)
 (\omega \tau_\mathrm{c}^{-1})^{1/2}
 \sim \sqrt{\omega \tau_\mathrm{conv}^{-1}},
\end{equation}
in terms of the convective velocity $v_\mathrm{conv}$,
turnover time $\tau_\mathrm{conv}$, typical length scale $L$,
and angular velocity $\omega$, growth may initially be fast
on a time scale between $\tau_\mathrm{conv}$ and the rotation
period. However, the $\alpha$-$\Omega$-dynamo may transition to much slower growth well before reaching saturation
\citep{raynaud_20}. As for the MRI, estimating the saturation field strength for the $\alpha$-$\Omega$ dynamo is not trivial. Assuming complete conversion of the free energy in differential rotation into
magnetic field energy \citep{duncan_92} is likely to significantly overestimate actual field strengths. Findings from simulations will be discussed in the next section.

Recently, the Tayler-Spruit dynamo \citep{spruit_02}, which has long been studied as a field amplification mechanism in the theory of stellar evolution, has also been considered as relevant for proto-neutron stars \citep{barrere_22}.
Different from the $\alpha$-$\Omega$-dynamo, the generation of a poloidal field from a toroidal field within the dynamo loop is mediated by the
Tayler instability \citep{tayler_73}, allowing the dynamo to operate in convectively stable regions as long as there is differential rotation to generate toroidal from poloidal fields.

Non-ideal effects present an important complication
both for convective dynamos in the proto-neutron star 
\citep{thompson_93} and for the MRI \citep{guilet_15}. While viscosity, heat conduction, and resistivity due to nucleons, electron, positrons, and photons are extremely small, non-ideal effects due to neutrinos play a major role, which can take on
the guise of a viscosity well above the typical neutrino mean free path or of drag below the neutrino mean free path. This leads to the possibility of a double cascade in proto-neutron star turbulence \citep{thompson_93} with potentially interesting ramifications for magnetohydrodynamic instabilities. 
The interplay of fermion chirality with magnetic fields and its potential effect in core-collapse supernovae has also been discussed
as an additional complication in the framework of chiral MHD \citep{masada_18}.

\subsection{Simulating Field Amplification Processes in Core-Collapse Supernovae}
\label{sec:amplification}
Due to the extreme resolution requirements for resolving the fastest growing MRI modes, studying field amplification in the context of magnetorotational supernovae is currently still a somewhat distinct endeavor from simulating magnetorotational explosions proper. Simulations focusing on field amplification processes need to sacrifice aspects that are relevant for modeling explosions, whereas explosion modeling (Section~\ref{sec:mrsims}) tends to circumvent thorny issues concerning field amplification by using relatively strong parameterized initial fields.
There is a significant history of local and semi-global simulations for studying field amplification by the MRI in the supernova context
\citep{obergaulinger_09,masada_12,guilet_15b,rembiasz_16b,guilet_22}.

In the last decade, simulations have begun to bridge the gap from MRI field amplification on small scales to the global scale. A first attempt was made using the thin layer approximation as a compromise to achieve sufficiently fine resolution 
\citep{masada_15}, which, however, does not permit conclusions about the emerging global field geometry. First truly global general relativstic MHD simulations
of MRI and dynamo field amplification in proto-neutron stars indeed showed rapid exponential field amplification to peak field strength well above
$10^{15}\,\mathrm{G}$, as well as the emergence of a global field due to the inverse cascade \citep{moesta_15}. These simulations achieved resolutions as fine
as $50\,\mathrm{m}$ and even included a basic treatment of neutrinos by means of a leakage scheme. However, they remained limited to very short simulations times of order $10\,\mathrm{ms}$, and still started from relatively strong initial field strengths. Subsequent simulations have shed more light on the long-term evolution of field amplification by using pseudo-spectral methods for high accuracy and avoiding fully compressible MHD. 3D MRI simulations over
more than half a second using incompressible MHD for a somewhat idealized proto-neutron star model
confirmed the emergence of such high field strength, but with modest energy in the dipole,
and, importantly, misalignment of the dipole from the axis \citep{reboul_21}.

More recent long-time pseudo-spectral simulations
with the \textsc{MagIC} code
have added realism by accounting for buoyancy effects with the Boussinesq approximation,
starting from PNS structures inspired by 1D core-collapse supernova simulations, and implementing
physical approximations for (neutrino) viscosity, thermal conductivity and magnetic diffusivity
\citep{raynaud_20,reboul_22,barrere_23}.
Such simulations of convection in rapidly rotating proto-neutron stars have demonstrated the generation of
strong dipole fields by the $\alpha$-$\Omega$ dynamo of order $10^{15}\,\mathrm{G}$ and even higher maximum toroidal field strengths of
order $10^{16}\,\mathrm{G}$, and also established scalings with proto-neutron star rotation period and Rossby number \citep{raynaud_20}.
Importantly, 
the initial exponential growth of the $\alpha$-$\Omega$ dynamo in these simulations stagnates at a first plateau and is then followed by very slow, secular amplification until saturation is reached only after several seconds. 
More recently, the emergence of large-scale fields
generated by an $\alpha$-$\Omega$ dynamo has also
been studied in compressible MHD simulations of proto-neutron star convection over shorter time scales \citep{masada_22}, confirming a trend towards stronger
large-scale fields at higher spin rate.

Subsequent simulations
with \textsc{MagIC}  have also addressed the convectively stable ``atmosphere'' at the proto-neutron star surface \citep{reboul_22}, which would seem to be more directly relevant to eventual surface field strengths and hence late-time powering of magnetized outflows, barring the possibility of field breakout from the interior. The simulations of \citet{reboul_22} , which cover $\gtrsim 1\,\mathrm{s}$ of field evolution, showed magnetar-grade fields of $\sim 10^{14}\,\mathrm{G}$ locally, but a relatively weak (and tilted) dipole with of $\lesssim 10^{13}\,\mathrm{G}$. Their simulations also show characteristic dynamo cycles with periods of several hundred milliseconds.

It has been theorized that field amplification in young neutron stars may also occur much later due
to fallback that spins up the outer layers of the neutron star
\citep{barrere_22}. This scenario presupposes that an explosion has already occurred by other means and is thus not directly relevant for the magnetorotational mechanism proper, but such late-time field amplification is relevant for magnetar formation and the possibility of increasing the energy or brightness of a supernova by means of an ``afterburner''. Since this scenario would result in a positive angular velocity gradient (faster surface rotation), MRI would not act in this case, but amplification by the Tayler-Spruit dynamo will be relevant \citep{barrere_22}. The 3D simulations of \citet{barrere_23} investigated this scenario for various degrees of differential rotation in the proto-neutron star, identifying distinct branches of the Tayler-Spruit dynamo with scaling laws for the saturation fields conforming to the theories
of \citet{spruit_02} and \citet{fuller_19}, respectively. The dynamo-generated fields are concentrated near the rotation axis, and dipole field strengths of $10^{14}\,\mathrm{G}$ or more can be reached.

\subsection{Simulations of Magnetorotational Explosions}
\label{sec:mrsims}
Global simulations of magnetorotational explosions date back to the 1970s. Even though many early numerical models
attempted to scale the problem down to one dimension \citep{bisnovatyi_76,mueller_79}, early work
by \citet{leblanc_70} already attempted to tackle it in 2D
on a $40\times 40$ grid, even with some very crude gray treatment of neutrino transport -- a pioneering approach despite the obvious limitations of the time.

\subsubsection{Prelude: Axisymmetric Simulations}
Modern 2D simulations started to revisit the problem of magnetorotational explosions about twenty years ago with mature numerical methods for MHD. In such 2D simulations, strong initial fields of order $10^{12}\,\mathrm{G}$ need to be prescribed by hand in order
for magnetic fields to influence the dynamics of the post-bounce evolution even if some amplification by field winding and 2D instabilities may occur. Early modern 2D simulations initially started with simplifications to the microphysics and neutrino transport to various degrees, but were able to model the extraction of rotational energy by magnetic fields and its impact on the shock dynamics, including the emergence of bipolar explosions
\citep{yamada_04,ardeljan_05,sawai_05,obergaulinger_06a,obergaulinger_06b,moiseenko_06,sawai_08,takiwaki_09,sawai_13,sawai_14,sawai_16}. Some works also investigated the repercussions of the specific dynamics of asymmetric collapse and explosion in the magnetorotational paradigm on neutrino and gravitational wave emission
\citep{kotake_04,obergaulinger_06a,takiwaki_11}.

A major breakthrough came in the form of the first
MHD supernova simulations \citep{dessart_07a} with multi-group neutrino transport
with the \textsc{Vulcan} code \citep{livne_07}, albeit using Newtonian gravity with no relativistic corrections. Even in the magnetorotational scenario, proper inclusion of neutrinos in simulations is of paramount importance as neutrino transport affects the proto-neutron star structure and contraction, will have some dynamical effect (even if it is subdominant to that of magnetic fields), and is critical for accurate predictions of nucleosynthesis from magnetorotational explosions.
For initial fields of order $10^{11}\texttt{-}10^{12}\,\mathrm{G}$, the simulations of \citet{dessart_07a} showed the emergence of bipolar magnetic tower structures that push the shock out and produce explosions with energies of up to more than $10^{51}\,\mathrm{erg}$, with some ejecta reaching velocities of $\sim 50,000\,\mathrm{km}$. They suggested that balance between spin-up of the proto-neutron star surface and  extraction of free rotational  by the jets can be maintained for time scales of seconds to eventually reach hypernova energies. Importantly, the jets remain in the non-relativistic regime, however, and are to be distinguished from relativistic long gamma-ray burst jets that could only emerge later in a more dilute environment to reduce baryon loading. A host of further
2D MHD simulations with neutrino transport has confirmed this overall picture and explored dependencies on progenitor mass, progenitor rotation profiles, and initial magnetic fields
\citep{obergaulinger_17,obergaulinger_19,obergaulinger_22,jardine_22}.

As an interesting variation of the magnetorotational
mechanism, follow-up work with the \textsc{Vulcan} code
considered the magnetorotational explosions after the accretion-induced collapse (AIC) of white dwarfs \citep{dessart_07b}, a scenario where rapid progenitor rotation can be justified naturally. For strong initial fields of $\sim 10^{12}\,\mathrm{G}$ powerful
explosions of $\sim 10^{51}\,\mathrm{erg}$ (i.e., not in the hypernova regime) can emerge. Due to negligible production
of $^{56}\,\mathrm{Ni}$, such explosions would be expected to
be optically dark, but would be prodigious production sites of $r$-process material due to the ejection of very neutron-rich material
with electron fraction $Y_\mathrm{e}$ down to $\sim 0.1$, which would imply strong constraints on AIC event rates.

\subsubsection{Dynamics of Magnetorotational Explosions in 3D}
First ventures into 3D MHD simulations of rotational core collapse also started in the late 2000s, and initially relied on heavily simplified physics without a microphysical equation of state and neutrino effects \citep{mikami_08,kuroda_10}.
3D simulations with a microphysical equation of state, a parameterized
treatment of deleptonization during collapse, and  parameterized initial rotation profiles and magnetic fields
initially focused on the gravitational wave emission during the early post-bounce phase and allowed limited conclusions on explosion dynamics and outcomes
\citep{scheidegger_08,scheidegger_10}.
Subsequent 3D simulations with neutrino leakage 
and very strong poloidal initial fields of $5\times 10^{12}\,\mathrm{G}$
showed the emergence of jets as in 2D models and the ejection of neutron-rich material with 
potential for r-process nucleosynthesis \citep{winteler_12} (for a more detailed discussion of nucleosynthesis conditions see Section~\ref{sec:multi_messenger}). The first 3D  simulations of magnetorotational supernovae were, however, too short to investigate explosion energetics and could only follow the growth of the explosion energy to $\sim 10^{50}\,\mathrm{erg}$, well below the energies of normal supernovae, let alone hypernovae.

\begin{figure}
    \centering
    \includegraphics[width=0.49\linewidth]{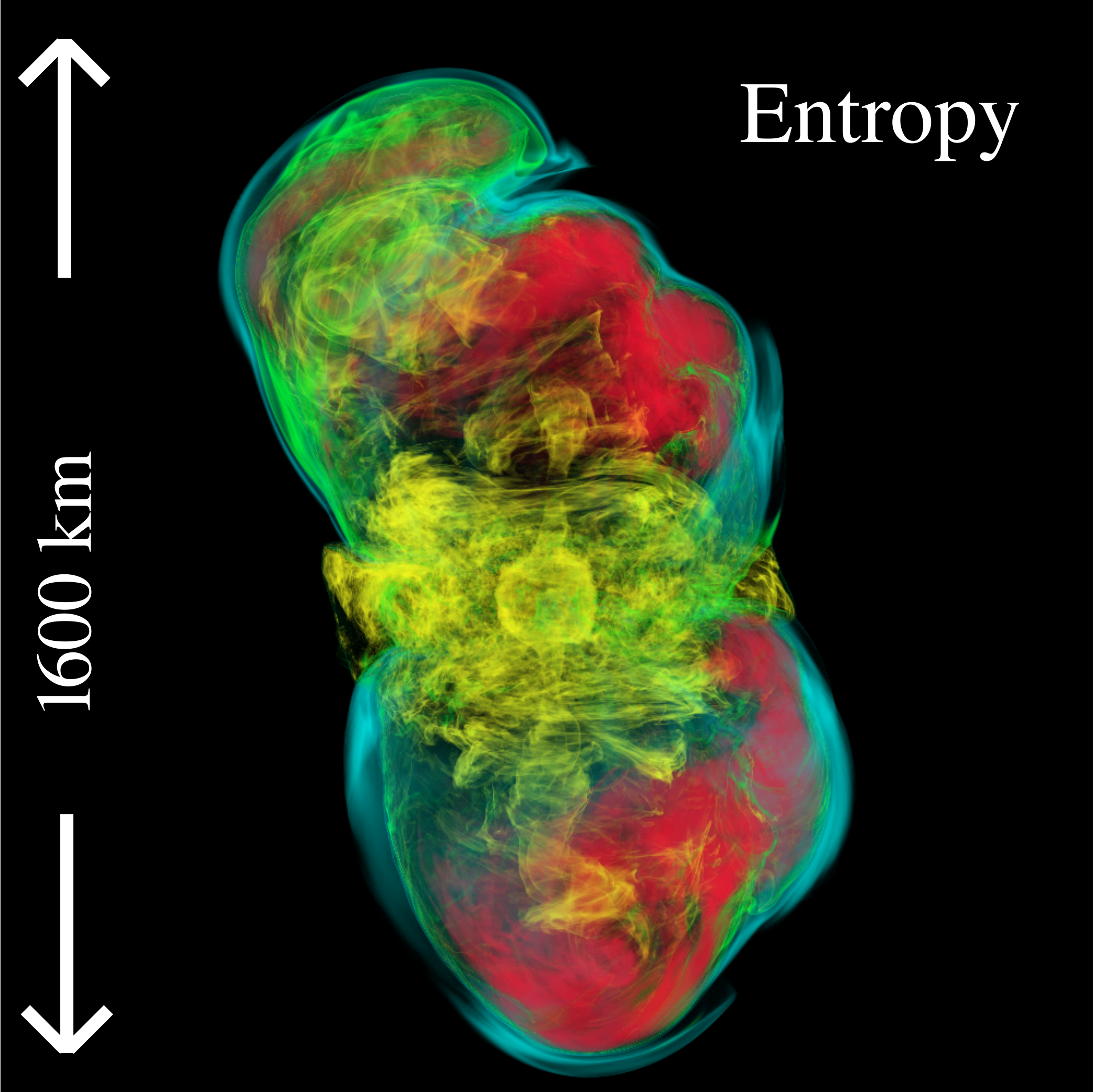}
    \includegraphics[width=0.49\linewidth]{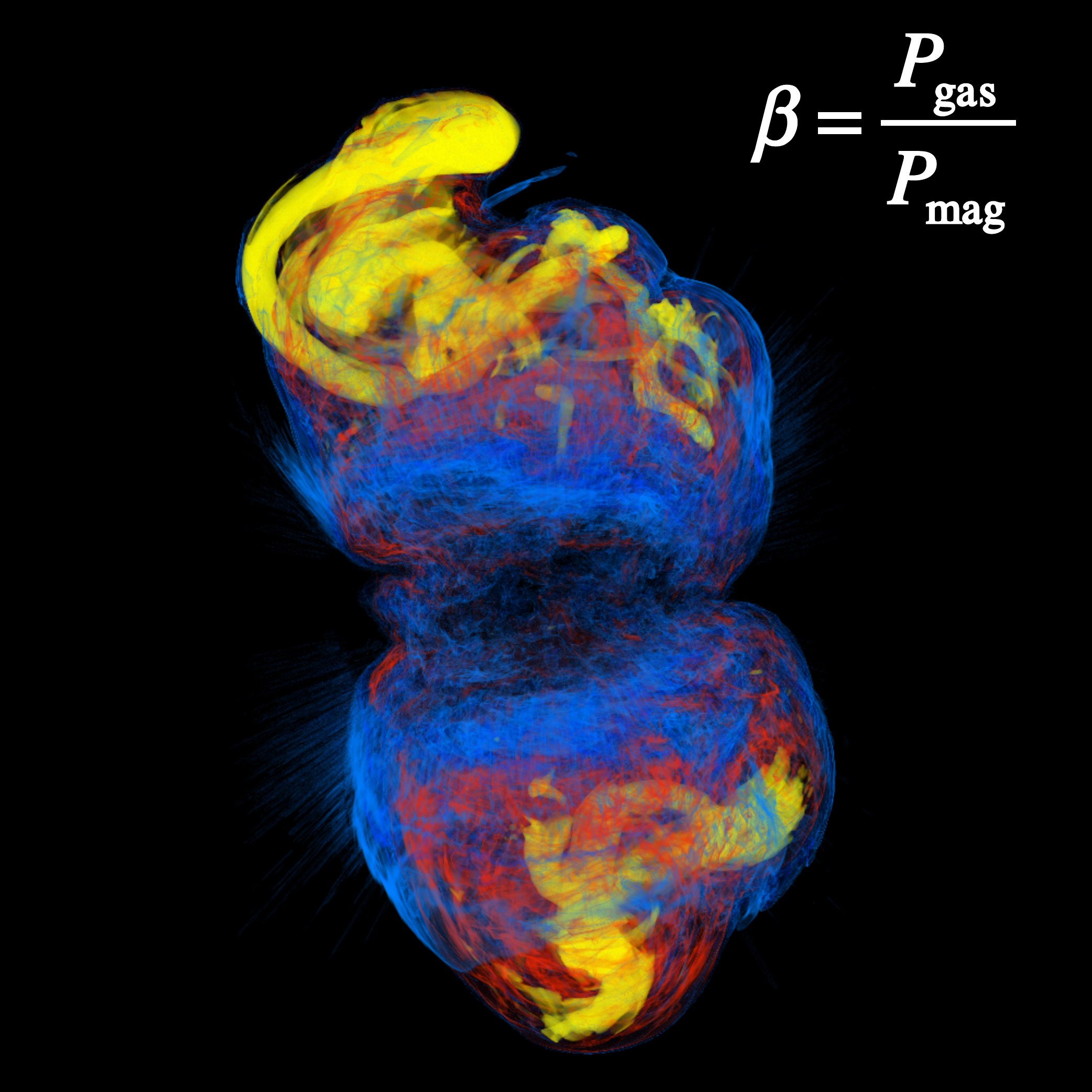}
    \caption{Volume rendering of entropy (left) and plasma-$\beta$ (right) in a simulation of a magnetorotational supernova \citep{moesta_14b} at a post-bounce time of $161\,\mathrm{ms}$. The initial jet has been disrupted by the kink instability, but strongly magnetized lobs (yellow) still continue to expand and have pushed the shock (cyan translucent surface in the left panel)  out to almost $1000\,\mathrm{km}$ in radius. 
    Figure from \citep{moesta_14b}.
    \href{
https://doi.org/10.1088/2041-8205/785/2/L29
}{DOI:10.1088/2041-8205/785/2/L29}.
    \textcopyright{AAS}.
    Reproduced with permission.}
    \label{fig:moesta}
\end{figure}

Longer 3D leakage simulations \citep{moesta_14b} 
again with purely poloidal, but slightly
weaker initial fields  of $10^{12}\,\mathrm{G}$
using the \textsc{GRHydro} code flagged an important complication in 3D in the form of the kink instability
of the emerging jet
(Figure~\ref{fig:moesta}). Kink instability for a magnetically confined plasma cylinder is expected to occur for
wavelengths $\lambda$ \citep{kruskal_58,begelmann_98,moesta_14b}, 
\begin{equation}
    \lambda
    \gtrsim \frac{2\pi a}{
    \langle B_\mathrm{tor}/B_\mathrm{pol}\rangle},
\end{equation}
in terms of the toroidal and poloidal field strengths
$B_\mathrm{tor}$ and $B_\mathrm{pol}$ and
the cylinder radius $a$.  Strong toroidal fields that precipitate kink instability emerge unavoidably due to field winding and are a prerequisite to jet launching in the first place. Therefore kink instability at some wavelength is expected quite generically, though the growth time scale must be sufficiently short in relation to the propagation of the jet to affect it.
\citet{moesta_14b} find
that an $m=1$ mode immediately disrupts the forming jet
in full 3D, different from a control simulation in 
octant symmetry that does not allow for an $m=1$
instability.  Magnetic lobes still push the shock
out to almost $1000\,\mathrm{km}$ by the end of the simulation $185\,\mathrm{ms}$ after bounce, but accretion onto the proto-neutron star is not quenched.
\citet{moesta_14b} have argued that the phenomenon of jet disruption should be quite generic, especially given that jet launching is likely to occur on longer time scales if the pre-collapse magnetic fields first need to be amplified to launch jets, and hence suggested that the ultimate fate of such ``fizzling'' magnetorotational explosions was black hole formation and potentially a hypernova driven by a collapsar engine on longer time scales. General relativistic MHD simulations of magnetorotational explosions
with multi-group neutrino transport 
with the \textsc{fGR1} code revealed somewhat similar phenomenology where the jet is disrupted and only one big strongly magnetized lobe survives \citep{kuroda_20}. However, neutrino heating helps support the developing explosion, and a modest explosion energy of $\sim 10^{50}\,\mathrm{erg}$ is obtained in \citep{kuroda_20}, which still falls far short of the hypernova regime. More stable jets and an explosion energy
of up to $6\times 10^{50}\,\mathrm{erg}$ were found
in subsequent simulations with the same code
in a small parameter study with different initial rotation
rates and initial fields of $10^{12}\,\mathrm{G}$ and above \citep{shibagaki_23}.

Subsequent work has, however, drawn a more nuanced picture of the disruptive role of the kink instability.
\citet{halevi_18} showed that with an \emph{extremely}
strong initial poloidal field of $10^{13}\,\mathrm{G}$, the jet can be stabilized against the kink instability, even in the case of a magnetic dipole field somewhat misaligned with the rotation axis. The 3D simulations of
\citet{obergaulinger_21} set another optimistic counterpoint, finding the emergence of stable jets
with only mild kink instability with more modest
initial fields as low as $\sim 10^{11}\,\mathrm{G}$
with mixed poloidal and toroidal geometry in a $39 M_\odot$ progenitor \citep{aguilera_18}. Interestingly, they found a jet-driven magnetorotational explosion even for the \emph{predicted} field strengths in the underlying stellar evolution model (though this initial
field remains far from self-consistent due to assumptions
about field geometry). Due to long physical simulation times, \citet{obergaulinger_21} are able to reach  explosion energies of $\sim 10^{51}\,\mathrm{erg}$ and even $\sim 10^{52}\,\mathrm{erg}$ for initial fields
of  $\sim 10^{12}\,\mathrm{G}$, lending more credence to the magnetorotational mechanism as a possible scenario for hypernovae. Low radial (300 zones) and angular resolution ($2.8^\circ$) is a caveat, however, as this could potentially inhibit the growth of the kink instability. Higher-resolution 3D simulations of the same $39 M_\odot$ progenitor by \citet{powell_23} with mixed poloidal/toroidal fields of $10^{10}\,\mathrm{G}$
and $10^{12}\,\mathrm{G}$ also indicate that hypernova energies can be
achieved in magnetorotational explosions. However, their explosion models still differ from the familiar picture of jet-driven magnetorotational explosions. In their models, most of the rotational energy of the proto-neutron star is transferred by magnetic stresses into relatively spherical outflows with quick spin-down of the neutron star by about $250\,\mathrm{ms}$ after bounce. Jets emerge as an epiphenomenon and carry only a small fraction of the explosion energy. The jets are subject to kink instability to varying degree with more stable jets in the $10^{10}\,\mathrm{G}$ model and episodic disruption
and reemergence of the jets in the $10^{12}\,\mathrm{G}$ model. The explosion energies remain at the lower end of
the hypernova range with $3\times 10^{51}\,\mathrm{erg}$.

The above panorama of recent simulations of magnetorotational explosions is ambiguous as there are significantly different outcomes in models by different groups. This is partly due to different choices of progenitor models and initial conditions for the magnetic fields. But there is also a clear need for controlled code comparisons to gauge to what degree the different outcomes, in particular with respect to jet disruption by the kink instability, are due to numerical uncertainties. 

Furthermore, the role of the \emph{initial} pre-collapse fields is increasingly recognized as a crucial missing piece in the magnetorotational paradigm, regardless of whether significant amplification of field \emph{strengths} is required after collapse or not. The default assumption of poloidal or poloidal-toroidal dipole fields aligned to the rotation axis in most global simulations constitutes a strong assumption and is not in line with the phenomenology of convective stellar dynamos or expectations for the Tayler-Spruit dynamo. Simulations with inclined magnetic dipoles \citep{halevi_18,bugli_21}
and quadrupolar pre-collapse fields \citep{bugli_21} suggest a significant impact on the initial field geometry on the post-collapse dynamics. Thus, similar to
the case of no or slow progenitor rotation, there is a need for 3D simulations of the pre-collapse phase
(Section~\ref{sec:precollapse}).

Compared to non-rotating or slowly
rotating massive stars, MHD simulations
of hypernova progenitors present significantly bigger challenges, however.
Most importantly, such simulations need to rely on rotation profiles from 1D stellar evolution models, and any deviations from the ``true'' quasi-equilibrium angular momentum distribution will trigger significant numerical transients. Although one must bear these caveats in mind, first 3D MHD simulations of shell burning in rapidly rotating massive stars have been performed and yield some interesting insights on hypernova
progenitor evolution. In an MHD
simulation of a $16 M_\odot$ progenitor,
\citet{varma_23b} find the development
of an $\alpha$-$\Omega$ dynamo that
generates strong fields of more than $10^{10}\,\mathrm{G}$ in the oxygen and neon shell, but the dipole component remains about an order of magnitude weaker. Importantly, such fields cannot be directly compared to the pre-collapse \emph{core} field strengths quoted for typical 3D MHD supernova simulations; the usual dipole fields in 3D MHD supernova models would have significantly weaker fields in the oxygen shell than in the core. However, the work of \citep{varma_23b} also highlights the inconsistencies of the current treatment of angular momentum transport and magnetic fields in 1D stellar evolution. In their 3D model, the dynamo-generated fields become strong enough to rapidly redistribute angular momentum between shells and suppress convection in the oxygen and neon shell. Efficient rotational coupling with the shells further out spins down the oxygen shell
from an angular velocity of
$0.07 \,\mathrm{rad}\,\mathrm{s}^{-1}$
initially to about
$0.01 \,\mathrm{rad}\,\mathrm{s}^{-1}$
after $500\,\mathrm{s}$ of simulation time.
While much work is needed to investigate the role of resolution, the possible role of the magnetic Prandtl number, and evolution over longer time scales, the findings of \citep{varma_23b} highlight that our incomplete understanding of angular momentum transport and magnetic field generation in stellar evolution models is a key uncertainty for the magnetorotational explosion scenario.

\subsubsection{Multi-Messenger Signatures from Magnetorotational Explosions}
\label{sec:multi_messenger}
A principal challenge in determining the explosion mechanism of hypernova lies in identifying suitable direct or indirect observables that can potentially discriminate between different scenarios or at least define ``litmus tests'' for simulations. Unlike for normal supernovae, compact remnant populations provide little opportunities to constrain the nature of hypernova explosions, as these make up only for a small fraction of compact object formation events. This leaves the transients themselves and their contribution to the chemical evolution of galaxies as the key sources for direct and indirect clue on the hypernova mechanism.
Emission from the long gamma-ray bursts often associated with hypernovae has received significant attention, but as the relativistic gamma-ray burst (GRB) only contains a minor fraction of the total explosion energy \citep{woosley_06b}, it does not provide direct clues to the hypernova explosion mechanism. Furthermore, the features of the GRB and its X-ray
afterglow are often ambiguous; for example, plateaus
in  X-ray afterglows have been taken as evidence for
millisecond magnetars \citep{corsi_09,gompertz_14}, but can in
fact also be accommodated in the collapsar model
\citep{duffell_15}. Lastly, there is still a gap between the core-collapse supernova simulations discussed in the this chapter in the previous subsection and models of the later GRB phase.

\begin{figure}
\centering
    \includegraphics[width=0.65\linewidth]{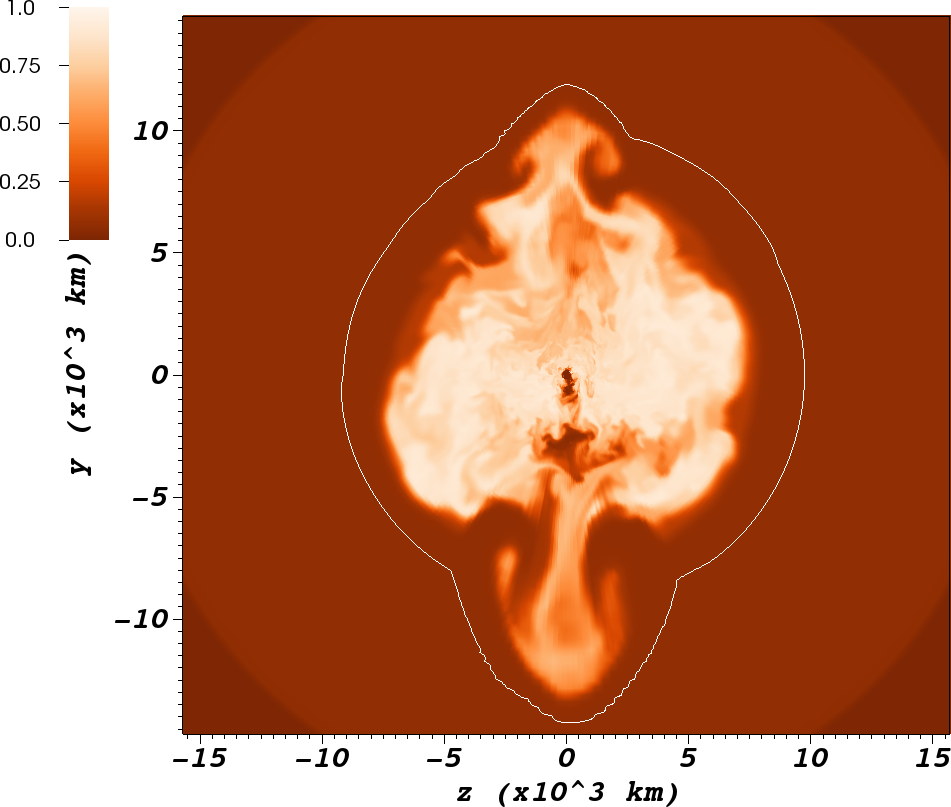}
    \\
    \includegraphics[width=0.65\linewidth]{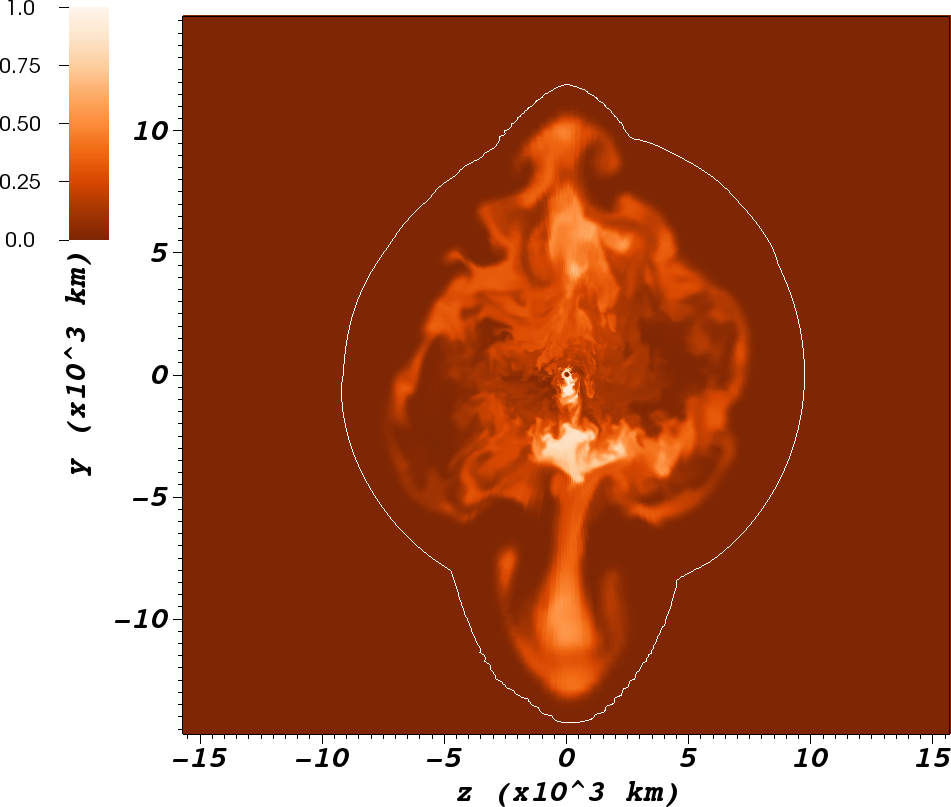}
    \caption{Mass fractions of
    iron-group elements (top)
    and $^4\mathrm{He}$ (bottom)
    meridional slice through the
    two magnerotoational explosion models of \citet{powell_23} with a pre-collapse
    core field strength of $10^{12}\,\mathrm{G}$ by the end of the simulations. 
    The shock is denoted by a white line.
    Note relatively spherical distribution of the iron-group ejecta, with only minor protrusion along the axis due to the penetrating jet. The jets have a bigger impact on shaping the distribution of $^4\mathrm{He}$, which is produced
    more abundantly in the jets due to freeze-out from nuclear statistical equilibrium at higher entropy.}
    \label{fig:ig_ejcta}
\end{figure}

Among potential direct and indirect observables from hypernova explosions, nucleosynthesis outcomes have already been studied extensively on the basis of MHD supernova simulations (for a broader review of
nucleosynthesis in jet-associated supernovae see
\citep{obergaulinger_23b}).
The critical idea that neutron-rich matter from close to the proto-neutron star can
maintain a low electron fraction $Y_\mathrm{e}$ 
if it is rapidly ejected in jets, and then provide conditions for r-process nucleosynthesis has already
been formulated based on 2D simulations
\citep{nishimura_06,dessart_07b,fujimoto_08,grimmett_21,reichert_21}.
However, 3D modeling and neutrino transport are required for nucleosynthesis predictions to accurately capture both the explosion dynamics and the neutrino-processing of the ejecta, which tends to drive them to more proton-rich conditions as they expand to lower densities. The first generation of 3D MHD simulations still relied on neutrino leakage and therefore could not provide reliable $Y_\mathrm{e}$ in the ejecta directly. Post-processing to approximately capture the effects of neutrino exposure on the ejecta is possible to remedy this to some degree, however.
The first nucleosynthesis calculations based on 3D MHD models with neutrino leakage suggested very neutron-rich ejecta
in the jets down to $Y_\mathrm{e}\sim 0.15$, which is sufficiently neutron-rich for an r-process up to the third peak
\citep{winteler_12}.
Subsequent simulations with leakage indicated, however,
that r-process nucleosynthesis up to the third-peak is sensitive not only to assumptions about neutrino exposure, but can also be thwarted by misalignment of the (assumed) initial magnetic dipole field and jet disruption by the kink instability \citep{halevi_18,moesta_18}.
3D models with neutrino transport 
\citep{reichert_23,reichert_24,zha_24}
still exhbit ejection of neutron-rich matter in the jets,
but have yet to show yields consistent at best with
solar r-process abundances up to the third d peak,
although low-level production of elements up to the third
peak has been found in some energetic explosion models.
In some cases, there is little production of elements beyond $Z=40$ even for relatively energetic explosions \citep{zha_24}. In the recent models of
\citet{zha_24}, the most overproduced isotopes are
rather those close to the $N=50$ shell closure
(specifically $^{82}\mathrm{Se}$, $^{83}\mathrm{Kr}$, and $^{76}\mathrm{Ge}$). High overproduction factors indicate that these yield patterns can at most represent $10\%$ of all hypernovae to avoid conflicts with constraints from chemogalactic evolution. The current state of nucleosynthesis predictions based on magnetorotational core-collapse supernova simulations is thus far from a firm identification of hypernovae as a robust r-process site, and could indicate that the dynamics of early, non-relativistic jets (provided that they are present and stable) is not yet captured correctly in current models.

Production of $^{56}\mathrm{Ni}$ is a further benchmark for magnetorotational supernova models. Several
$0.1 M_\odot$ of $^{56}\mathrm{Ni}$ are required
to account for the luminosities of broad-lined Ic supernovae, though it is not perfectly clear whether decay of $^{56}\mathrm{Ni}$ is the only power source for the light curves. 3D MHD supernova simulations
are able to reach such nickel masses \citep{reichert_23,powell_23,zha_24}.
In future, comparison to supernova
spectropolarimetry may provide stronger
constraints for model validation. Observed broad-lined
Ic supernovae appear to be characterized by a bipolar distribution of iron-group ejecta, moving at velocities up to more than
$10,000\, \mathrm{km}\, \mathrm{s}^{-1}$,
and a toroidal distribution of oxygen
\citep{mazzali_01,maeda_08,tanaka_17}.
Such an ejecta structure is not yet seen in the explosion models of \citet{powell_23}; iron-group elements are instead distributed quite spherically and move at slower
speeds (Figure~\ref{fig:ig_ejcta}). However, the ejecta geometry can change substantially from the first second of
the explosion to shock breakout due to mixing instabilities. Long-time simulations beyond shock breakout are required and need to be fed into radiative transfer calculations for meaningful comparisons with observed transients. Pipelines from multi-D core-collapse supernova simulations to synthetic light curves and spectra are now becoming available \citep{maunder_24} and
will enable more stringent validation of magnetorotational supernova models in future.

In future, gravitational waves could serve as a more direct probe of the dynamics of hypernova explosions.
Observations of hypernovae with gravitational waves will, however, only be possible for nearby events even with third-generation detectors, and face the additional obstacle that the hypernova fraction is low in the local Universe. Nonetheless, gravitational wave signal predictions for magnetorotational explosions have been studied in some depth for clues about the explosion mechanism in such a fortuitous event. The bounce signal from the collapse of rapidly rotating cores is sometimes implicitly taken as a potential giveaway for magnetorotational explosions, but though it is a suitable feature for distinguishing the gravitational wave signal
from potential hypernova progenitors from ordinary core-collapse supernovae \citep{heng_09,logue_12,abdikamalov_14,powell_16}, it only probes the \emph{preconditions} for magnetorotational explosions rather than the dynamics of the explosion mechanism itself. After several earlier studies had already investigated the impact of MHD effects on the bounce signal with simplified microphysics
\citep{obergaulinger_06a,scheidegger_08,takiwaki_11},
gravitational wave signal predictions covering hundreds of milliseconds of the post-bounce and explosion phase have recently become available from MHD supernova simulations with multi-group neutrino transport
\citep{jardine_22,powell_23,powell_24}.
The characteristic imprint of strong, dynamically important magnetic fields in supernova explosions remains subtle, however. The post-bounce gravitational wave signal of magnetorotational explosions does not differ fundamentally from the standard time-frequency structure
of gravitational waves from core-collapse supernovae
\citep{mueller_13,abdikamalov_22} with a dominant rising
emission band at several $100\,\mathrm{Hz}$ to more
than $1\,\mathrm{kHz}$
from a low-order buoyancy-driven quadrupole oscillation
(f-mode or g-mode) living near the proto-neutron star surface and other less prominent features. The gravitational wave signals from magnetorotational
explosions differ from neutrino-driven ones in detail, though. Axisymmetric simulations of \citet{jardine_22}
suggested that magnetorotational explosions may be characterized by earlier subsidence of the high-frequency f/g-mode emission due to the quenching of accretion, and possibly a broader, more diffuse f/g-mode emission band.
Magnetorotational explosions with strong jets should also
exhibit a very strong tail signal due to the pronounced bipolarity of the expanding shock, but the detection of such a signal feature is not trivial because of its low-frequency nature. 3D simulations both with the
\textsc{CoCoNuT-FMT} and \textsc{Alcar} code
\citep{powell_23,powell_24} paint a similar picture, with the exception that no strong tail signal may emerge because the jets are weaker than in axisymmetry (even if they are not disrupted by the kink instability). 
Nonetheless, modern signal classification methods may be
able to detect the subtle differences in the time-frequency structure to distinguish magnetorotational
explosions from neutrino-driven ones at
moderately high overall signal-to-noise ratios of
$25\texttt{-}45$ \citep{powell_24}.
It is also noteworthy, however, that predicted gravitational wave signals from hypernovae are generally quite strong and detectable out to 
$2\texttt{-}5\,\mathrm{Mpc}$ with third-generation
detectors \citep{powell_24}.
Finally, another interesting prospect is that \emph{late-phase} gravitational wave emission may contain quite direct clues for dynamo-generated field in rapidly-rotating proto-neutron stars. In the case of rapid rotation, gravitational wave emission from the proto-neutron star convection zone is characterized by a spectrum with prominent, well-defined inertial mode
frequencies, whose amplitudes and frequency structure can be related to the rotation and magnetization of the proto-neutron star \citep{raynaud_22}. The detectability of such a signal remains to be investigated, however.

\section{Outlook}
Magnetohydrodynamic simulations of core-collapse supernovae have progressed considerably in recent years. Simulations combining magnetohydrodynamics and detailed neutrino-transport have become much more common, and are becoming important beyond the traditional ``niche'' of the magnetorotational explosion paradigm, especially for understanding neutron star birth spins and magnetic fields. Magnetorotational explosion models are becoming sufficiently mature to deliver better predictions for observables and sufficiently widespread to enable better characterization of uncertainties in the simulations and in predicted observables. But much remains to be done. Capabilities for end-to-end modeling need to be developed to connect core-collapse simulations proper with models for GRB launching from millisecond magnetars \citep{komissarov_07,bucciantini_08,bucciantini_09} and collapsar-driven explosions and jets \citep{barkov_08,fujimoto_08,ono_12,siegel_19,gottlieb_22a,gottlieb_22b}. Bridging the gap to neutron star birth magnetic fields also remains a challenge. Validation of MHD-driven supernova models by comparison to transient observations will become an important priority, and there is the need for a more integrated modeling of supernova progenitor evolution in 1D and 3D to address uncertainties in the interior rotation and magnetic field structure of massive stars. Magnetohydrodynamic supernova simulations are bound to develop into central hub in an interdisciplinary research network connecting computational astrophysics, stellar evolution, compact objects, and transient observations.

\begin{acknowledgement}
I wish to thank B.~Sykes for critical reading,
P.~M\"osta
for his kind permission to reproduce figures,
and J.~Powell for discussions and permission to reuse simulation data
for Figure~\ref{fig:ig_ejcta}.
I acknowledge support from the Australian Research Council through Future Fellowship FT160100035, as well as computer time allocations from Astronomy Australia Limited's ASTAC scheme, the National Computational Merit Allocation Scheme (NCMAS), and from an Australasian Leadership Computing Grant. Some of this work was performed on the Gadi supercomputer with the assistance of resources and services from the National Computational Infrastructure (NCI), which is supported by the Australian Government, and through support by an Australasian Leadership Computing Grant. Some of this work was performed on the OzSTAR national facility at Swinburne University of Technology. OzSTAR is funded by Swinburne University of Technology and the National Collaborative Research Infrastructure Strategy (NCRIS).
\end{acknowledgement}

\bibliography{paper}

\end{document}